\documentclass{elsart}

\usepackage{harvard}
\usepackage{epsfig}

\usepackage{amssymb}

\def\astrobj#1{#1}
\def\url#1{{\ttfamily\def\/{/\discretionary{}{}{}}#1}}
\def\bibcode#1{(\texttt{#1})}
\def\LCDM{$\Lambda${CDM}\ }
\begin{document}

\begin{frontmatter}
\title{ The little galaxy that could: Kinematics of Camelopardalis~B}
\author{Ayesha Begum\thanksref{email1}}
\address{NCRA/TIFR, P. O. Bag 3, Ganeshkhind, Pune 411007, India.}
\author{Jayaram N. Chengalur\thanksref{email2}}
\address{NCRA/TIFR, P. O. Bag 3, Ganeshkhind, Pune 411007, India.}
\author{Ulrich Hopp\thanksref{email3}}
\address{ Universitatssternwarte Munchen, Scheinerstrasse 1, 
	  D-81679 Munchen, Germany.}
\thanks[email1]{ayesha@ncra.tifr.res.in}
\thanks[email2]{chengalu@ncra.tifr.res.in}
\thanks[email3]{hopp@usm.uni-muenchen.de}

\begin{abstract}
	
	We present deep, high velocity resolution ($\sim 1.6$ km sec$^{-1}$)
Giant Meterwave Radio Telescope HI 21cm synthesis images, as well as
optical broad band images, for the faint ($M_B \sim -10.9$) dwarf 
irregular galaxy Camelopardalis~B.

	We find that the HI in the galaxy has a regular velocity
field, consistent with rotational motion. Further, the implied 
kinematical major axis is well aligned with the major axis of both
the HI flux distribution as well as that of the optical emission. 
Camelopardalis~B is the faintest known galaxy with such relatively
well behaved kinematics.

	From the HI velocity field we derive a rotation curve for
the galaxy using the usual tilted ring model. The rotation curve 
can be measured out to galacto-centric distances $> 4$ times the 
optical scale length. The peak (inclination corrected) rotation 
velocity $v_o$ is only $\sim 7$~km~sec$^{-1}$ -- the high velocity
resolution of our observations were hence critical to measuring
the rotation curve. Further, the peak rotational velocity is 
comparable to the random velocity $\sigma$  of the gas, i.e. 
$v_o/\sigma \sim 1$. This makes it crucial to correct the observed 
rotation velocities for random motions before trying to use the 
kinematics to construct mass models for the galaxy. After applying
this correction we find a corrected peak rotation velocity of
$\sim 20$~km~sec$^{-1}$.

	On fitting mass models to the corrected rotation curve 
we find that the kinematics of Camelopardalis~B can be well fit with 
a modified isothermal halo with central density 
$\rho_0 \sim 12$ $M_\odot$ pc$^{-3}$. This central density is 
well determined, i.e. it has a very weak dependence on the assumed 
mass to light ratio of the stellar disk. We also find that the 
corrected rotation curve cannot be fit with an NFW halo regardless 
of the assumed mass to light ratio.

	Finally we compile from the literature a sample of galaxies
(ranging from normal spirals to faint dwarfs) with rotation curves obtained
from HI synthesis observations. The complete sample covers a luminosity
range of $\sim 12$ magnitudes. From this sample we find (i)~that 
Camelopardalis~B lies on the Tully-Fisher relation defined by these 
galaxies, provided we use the pressure support  corrected rotation velocity, and 
(ii)~a weak trend for increasing halo central density with decreasing
galaxy size. Such a trend is expected in hierarchical models of halo
formation.

\end{abstract}

\end{frontmatter}

\section{Introduction}
\label{sec:intro}

	There are a number of reasons why kinematical studies of faint
dwarf galaxies are particularly interesting. The first is related to the 
structure of the dark matter halos of galaxies. Traditionally, dark 
matter halos have been modeled as ``modified isothermal'' halos 
for which the density is 
constant in the central regions. However, numerical simulations of 
galaxy formation in hierarchical structure formation models (such as 
the standard  CDM or \LCDM  models) predict that galaxy halos should 
have cuspy central density distributions \cite{navarro96,klypin01}.
In particular, for the popular \citeasnoun{navarro96} model, (generally 
referred to as the ``NFW'' halo), the density in the central regions 
increases as $r^{-1}$. Unfortunately, it turns out that it is difficult
to distinguish between isothermal and NFW halos from the observed
kinematics of normal spiral galaxies. This is because the gravitational 
force in the central parts of these galaxies is usually dominated by the
mass of the stellar disk. Disentangling the contribution from the dark
matter halo hence requires knowledge of the mass to light ratio (M/L) of
the stellar disk, which is often poorly constrained. Depending on the
assumed disk M/L ratio, equally good fits to the observed kinematics
can be obtained using either isothermal halos or NFW halos 
\citeaffixed{navarro98}{see eg.}. Dwarf galaxies, on the other hand, are 
often dark matter dominated, even in their inner parts, 
\citeaffixed{carignan88}{eg. DDO154,} and consequently their kinematics
can be used to constrain the structure of their dark matter halos with 
minimum uncertainties due to the unknown M/L ratio of the stellar disk. 

	Apart from the shapes of galaxy halos, numerical simulations
also predict correlations between various physical parameters of
the halo, most notably between the characteristic density and the virial
mass \citeaffixed{navarro97}{eg.}. The characteristic density is found to 
anti-correlate with the virial mass, i.e. the lowest mass halos have the 
highest densities. In hierarchical scenarios  the low mass halos form 
at the earliest times (at which time the background density is the highest)
and if there is a constant relation between the background density and 
the characteristic density of the collapsed halo then such a relation
is to be expected. The correlation found in numerical simulations is 
however both weak (a large change in the virial mass of the halo leads to
relatively small changes in the characteristic density) and noisy. To 
test whether such correlations are present in real galaxies it is 
useful to have data on as wide a range of galaxy masses as possible.
While there is already a wealth of kinematical data for bright spiral
galaxies, the number of dwarf galaxies, in particular extremely
faint (fainter than M$_B \sim -12$) dwarf galaxies, for which good 
kinematical data is available, is very limited.

	Finally, even from a purely phenomenological point of view, the 
kinematics of extremely faint dwarf irregular galaxies has been the subject 
of interest from some time now. This is because it is unclear whether the 
faintest dwarf irregular galaxies are rotationally supported or not. In a 
systematic study of 9 faint dwarf galaxies \citeasnoun{lo93} found that only
two galaxies showed ordered velocity fields, the remaining galaxies all 
had chaotic velocity fields. Further, independent observations of the
extremely faint dwarf galaxy GR8, \citeaffixed{carignan90}{M$_B\sim -11.0$,} 
showed that although the galaxy had a somewhat ordered HI velocity field,
the implied kinematical major axis was perpendicular to the major axis 
of the optical and HI disk. Similarly, \citeasnoun{cote00} found an ordered 
velocity field for SDIG (M$_B\approx-11.3$), but again the kinematical 
major axis was found to be perpendicular to the major axis of the optical
disk. All of this lead \citeasnoun{cote00} to suggest that normal rotation
is seen  only in dwarfs brighter than M$_B \sim -14$, and that by 
M$_B \sim -13$  one begins to find systems with misaligned axis, and 
other kinematical peculiarities. These peculiarities make it difficult,
if not impossible to model the kinematics of these galaxies using 
traditional mass models. This is unfortunate, since it is these very 
faint systems (which presumably formed at the  earliest epochs and 
are the most dark matter dominated) which could be particularly useful 
in determining dark halo properties.  

	We present here deep, high velocity resolution, Giant Meterwave Radio
Telescope (GMRT) observations of the faint (M$_B \sim -10.9$) dwarf irregular
galaxy Camelopardalis~B. We find that the galaxy has a very well ordered
velocity field, with no sign of major kinematical peculiarities, despite
the fact that the peak (inclination corrected) rotation velocity is
only $\sim 7$~km/s. The rest of this paper is divided as follows. Details of
the observational setup and procedure (both optical and HI 21cm) can be 
found in section~\ref{sec:obs}. The kinematics of the HI disk and derivation
of the rotation curve are presented in section~\ref{sec:rotcur}, mass models
for Cam~B are presented in section~\ref{sec:massmodels}. Finally in 
section~\ref{sec:dis} we discuss the kinematics of Cam~B in light of the 
various issues discussed above. Through out this paper we use a distance 
of 2.2~Mpc \citeaffixed{huchtmeier00}{given by } for Cam~B. 
Recent HST  observations (D.~J.~Bomans, private communication) confirm this
distance estimate.

\section{Observations}
\label{sec:obs}
\subsection{Optical Observations}
\label{ssec:opt_obs}

	CCD images of Cam~B were obtained in Oct. 1996 in the Johnson~B 
and V filters using the focal reducer system CAFOS at the Calar Alto~2.2m 
telescope. The instrument was equipped with a Tektronix 1024 by 1024 pixel
chip with one pixel corresponding to 0.53$^{''}$. The usual CCD calibration 
frames (bias, flats and blank fields) were obtained and standard fields 
from the list of \citeasnoun{landolt} and \citeasnoun{christian85} were 
observed in order to determine the absolute photometry. The exposure 
time on the target galaxy was 1200 and 700 seconds in B and V, respectively
and the FWHM seeing of the co-added images was 2.9~arc~sec. The total exposure
time was divided into three chunks and the pointing centers of the
individual exposures were shifted by several tens of arc seconds. 

\subsection{HI observations}
\label{ssec:gmrt_obs}
	
	The GMRT observations were conducted during the commissioning
phase of the telescope. The setup for the  observations is given  in 
Table~\ref{tab:gmrt}. Absolute flux calibration was done using scans on the 
standard calibrators 3C48 and  3C286 one of which was observed at the start
and end of each observing run. Phase calibration was done using 0410+769 
which was observed once every 30 minutes. Cam~B has V$_{\rm lsr} \sim
80$~km~sec$^{-1}$, and a velocity width of $\sim 20$~km~sec$^{-1}$, 
inspection of the Dwingeloo survey data toward Cam~B shows that no Galactic
emission is detected in this velocity range. Bandpass calibration was hence 
done in the standard way using 3C286 (which itself has no absorption 
features in the relevant velocity range).

     The data were reduced using standard tasks in classic AIPS.  For each run,
bad visibility points were edited out, after which the data were calibrated.
Calibrated data for all runs was combined using DBCON. The GMRT does not
do online doppler tracking -- any required doppler shifts have to be applied
during the offline analysis. However since the differential doppler shift 
over our observing interval is much less than the channel width, no 
correction needed to be applied offline. 

\begin{table*}
\caption{Parameters of the GMRT observations}
\label{tab:gmrt}
\vskip 0.1in
\begin{tabular}{ll}
\hline
Parameters& Value \\
\hline
\hline
RA(1950) & 04$^h$48$^m$03.3$^s$\\
Declination(1950) &  +${67}^{\circ} 01' 02''$\\
Central velocity (heliocentric) & 77.0 km sec$^{-1}$\\
Date of observations & 4$-$6 Nov 2001\\
Time on source & 16 hrs\\
Total bandwidth & 1.0 MHz\\
Number of channels & 128\\
Channel separation & 1.65 km sec$^{-1}$\\
FWHM of synthesized beam  & 40$^{''}\times38^{''}$, 24$^{''}\times22^{''}$, 16$^{''}\times14^{''}$\\
RMS noise per channel & 2.5~mJy, 2.0 mJy, 1.6~mJy\\
\hline
\end{tabular}
\end{table*}
 
\section{Results of the Observations }
\label{sec:res}
\subsection{Optical}
\label{ssec:opt_res}

	 Debiasing, flat-fielding and cosmic ray filtering of the CCD images
were done in the usual manner, using standard MIDAS routines. After this
the individual frames were added after shifting them to a common coordinate
system.
	
\begin{figure}[h!]
\begin{center}
\rotatebox{-90}{\epsfig{file=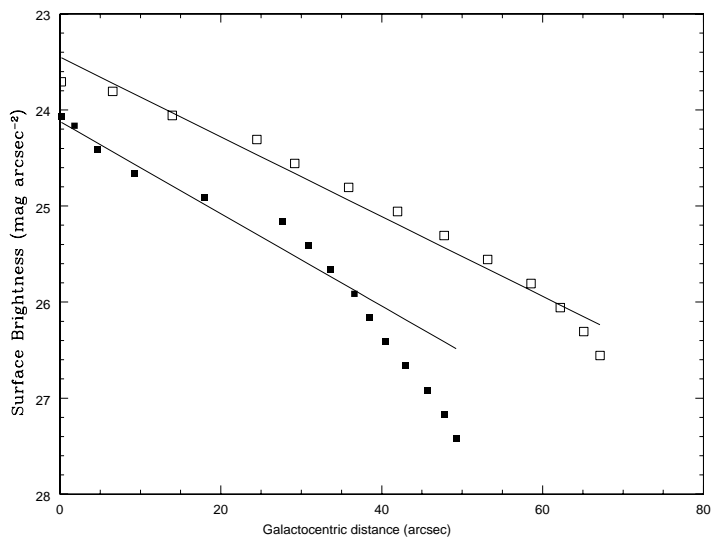,width=3in}}
\end{center}
\caption{Surface brightness profile of Cam~B in  the B (filled squares)
	and V (empty squares) bands. The best fit exponential profiles
	are shown superimposed.}
\label{fig:optprof}
\end{figure}

	Cam~B has a relatively irregular optical morphology 
(see Fig.~\ref{fig:overlay}). Because of the bad seeing, it was not possible
to determine whether the knotty structures seen in the B band image were 
barely resolved stars  or HII regions, consequently the distance to Cam~B
could not be estimated from this data set. The irregular morphology however 
means that the images needed to be smoothed before attempting surface 
photometry; a gaussian 10~arc~sec (FWHM) filter was used for this purpose. 
Surface photometry was done using the ellipse fitting algorithm of 
\citeasnoun{bender87}. The average ellipticity of Cam~B was found to 
be $0.53\pm0.16$ (corresponding to an inclination of $\sim 65^o$ for an
intrinsic thickness ratio $q_0 =0.25$, see section~\ref{sec:massmodels}).
The overall orientation of the opital major axis in the B (V) band data 
is 210$^o$ (199$^o$) degrees, in good agreement with the orientation of 
the HI data (see below). At the northen rim, a very faint,fuzzy extension
is visible in the V-band (and to a much smaller extend in the B-band) 
which changes the PA of the outermost isophotes towards a slightly lower 
value of 189$^o$, but still in good agreement with the overall distribution 
of the radio data. This extension is related to the color gradient seen 
in Fig.~\ref{fig:optprof}, which shows the surface brightness profiles 
in the B and V bands. Exponential fits to the surface brightness profiles
are shown superposed. The exponential scale lengths are $26.2^{''}$ for 
the V band and $22.6^{''}$  for the B band. The corresponding linear 
quantities (at an adopted distance of 2.2~Mpc) are 0.28~kpc and 0.24~kpc
respectively. From the figure it can also be seen
that $B-V$ is approximately constant in the inner regions of the galaxy 
(up to $\sim 35^{''}$), while a color  gradient is seen in the outer 
regions of the galaxy. Finally a mask at the 26.5 mag per square arc second
level was constructed in order to determine the flux in both filters within 
the Holmberg isophote.  The total B and V magnitudes within the 26.5 mag per square arc sec are found to be 16.71 and 15.91 respectively. The absolute Holmberg magnitude  obtained \citeaffixed{schlegel98}{after correcting for Galactic extinction using $A_B$=0.93~mag and $A_V$=0.72~mag,}, are  $M_B$= $-$10.94 and $M_V$=$-$11.52.  Note that no correction for internal extinction has been applied. The corresponding B and V luminosities are $L_B$=3.7$\times 10^6 L_\odot$ and $L_V$=3.5 $\times 10^6 L_\odot$. We also note that the low value of the
central surface brightness (Fig.~\ref{fig:optprof}) makes Cam~B an LSB
galaxy.

\subsection{ HI Flux and Surface Density Distribution }
\label{ssec:hi_flux}

     The GMRT has a hybrid configuration \citeaffixed{swarup91}{see}
with 14 of its 30 antennas located in a central compact array with 
size $\approx$ 1 km ($\approx$ 5 k$\lambda$ at 21cm) and  the remaining
antennas distributed in a roughly ``Y'' shaped configuration, giving a 
maximum baseline length of $\approx$ 25 km ($\approx$ 120 k$\lambda$ 
at 21 cm). The baselines obtained from antennas in the central square 
are similar in length to those of the ``D'' array of the VLA while 
the baselines between the arm antennas are comparable in length 
to the ``B'' array of the VLA. A single observation with the GMRT 
hence yields information on both large and small angular scales. Data 
cubes were therefore made at various (u,v) ranges, viz. 0$-$5 k$\lambda$, 
0$-$10 k$\lambda$ and  0$-15$ k$\lambda$ corresponding to  angular 
resolutions of 40$''\times38''$, 24$''\times22''$  and 16$''\times14''$ 
respectively. All the data cubes were deconvolved using the the AIPS 
task IMAGR. 

    The HI emission  from Cam~B  spanned  18 channels of  the
spectral cube. A continuum image was made using the average of all the
line free channels, no continuum was detected from the disk of 
Cam~B to a $3\sigma$ flux limit of 1.8~mJy per beam (for a beam 
size of $46^{''}\times37^{''}$). We also checked for the presense
of a few faint small continuum sources in the disk of Cam~B by 
making a high resolution ($3.6^{''}\times3.2^{''}$) map -- no sources
were detected down to a $3\sigma$ limit of 0.6~mJy.

\begin{figure}[h!]
\begin{center}
\rotatebox{-90}{\epsfig{file=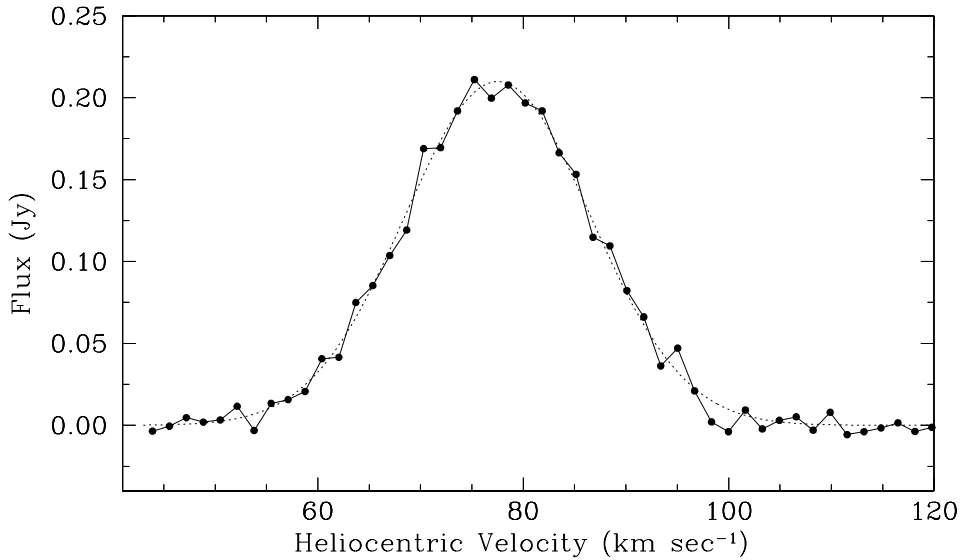,width=4in}}
\end{center}
\caption{ The global HI profile of Cam~B as derived from the GMRT data. The 
channel separation is $1.65$~km~sec$^{-1}$.  The dotted line shows a
Gaussian fit to the line profile.}
\label{fig:gprof}
\end{figure}

     The global HI emission profile of Cam~B is given in Fig~\ref{fig:gprof}.
A Gaussian fit to the profile gives a central velocity (heliocentric) of 
 $77.5\pm1.0$~km~sec$^{-1}$. The integrated flux is 
$4.6\pm0.4$~Jy~km~sec$^{-1}$. These are in excellent agreement with the
values of $77.0\pm1.0$~km~sec$^{-1}$ and  $4.47$~Jy~km~sec$^{-1}$ obtained
from single dish observations \cite{huchtmeier00}. The excellent agreement
between the GMRT flux and the single dish flux shows that the no flux was
missed because of the missing short spacings in the interferometric
observation. The velocity width at 50 \% level  of peak emission 
($\Delta V_{50}$) is found to be  $21.4\pm1.0$~km~sec$^{-1}$, which again
is in reasonable agreement with the $\Delta V_{50}$ value of $18$~km~sec$^{-1}$
determined from the single dish observations. The HI mass obtained 
from integrated profile (taking the  distance to the galaxy to be 
2.2~Mpc) is $5.3\pm0.5 \times{10}^{6} M_\odot$, and the $M_{HI}/L_B$
ratio is found to be $1.4$ in solar units.

	 We examined the line profiles at various locations in the galaxy 
and found that they were (within the noise) symmetric and  single peaked. 
Moment maps i.e. maps of the total integrated flux (moment~0), the flux 
weighted velocity (moment~1) and the flux weighted velocity dispersion 
(moment~2) were then made from the data cubes using  the AIPS task MOMNT. 
To obtain the moment maps, lines of sight with a low signal to noise ratio
were excluded by applying a mask at $3\sigma$ level, $\sigma$ being the rms
noise level in a line free channel. Maps of the velocity field and the
velocity dispersion were also made in GIPSY using gaussian fitting to the
individual profiles. As to be expected, given the simple line profiles, 
the AIPS moment~1 map and the velocity field obtained using gaussian fits
showed excellent agreement.  However the AIPS moment~2 map systematically 
underestimated the velocity dispersion (as obtained from gaussian fitting)
particularly near the edges where the signal to noise ratio is low. This can be
understood as the effect of the thresholding algorithm used by the
MOMNT task to identify the regions with signal. From the gaussian fitting
we find that the velocity dispersion $\sigma$ is $\approx$ 7.3 km sec$^{-1}$,
and shows no measurable variation across the galaxy. This value of $\sigma$
and the lack of measurable variation of $\sigma$ across the galaxy is 
typical of dwarf galaxies \citeaffixed{lake90,skillman}{eg.}. 

\begin{figure}[h!]
\begin{center}
\epsfig{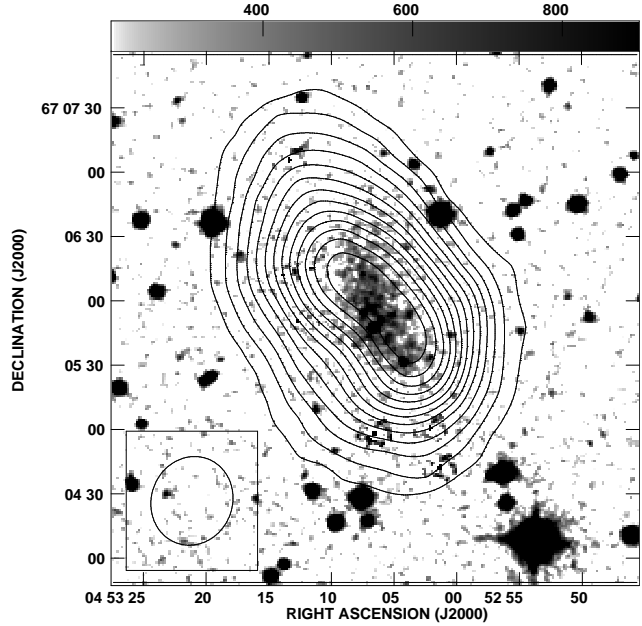}
\end{center}
\caption{The digitized Palomar Sky Survey image of Cam~B with the GMRT
	 $40^{''} \times 38^{''}$ resolution integrated HI emission 
	 (moment~0) map overlayed. The contour levels are $3.7,8.8,19.1,
	 24.3, 29.4,34.6,39.8,44.9,50.1,55.2,60.4, 65.5 \& 70.7 
	 \times 10^{19}$ atoms~cm$^{-2}$.
	}
\label{fig:overlay}
\end{figure}

\begin{figure}[h!]
\begin{center}
\rotatebox{-90}{\epsfig{file=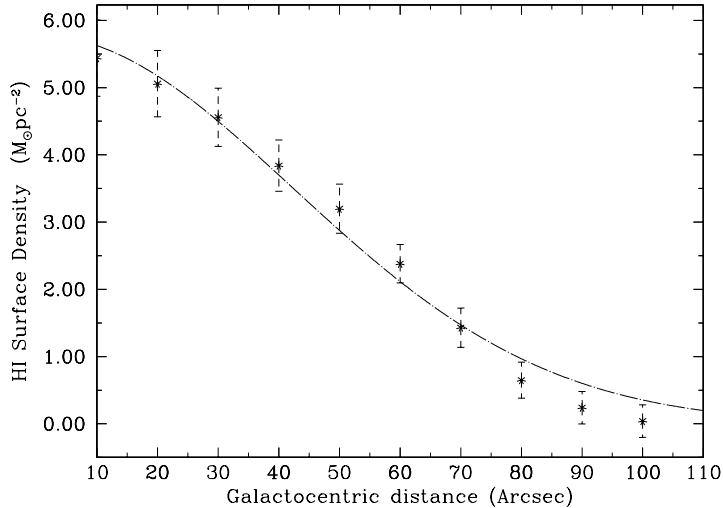,height=4.0in}}
\end{center}
\caption{The HI surface density profile derived from the HI surface
	 density shown in Figure 3. A gaussian fit is shown superimposed.
        }
\label{fig:hisb}
\end{figure}

    Figure~\ref{fig:overlay} shows the integrated HI emission at 
$40^{''}\times38^{''}$~arcsec resolution overlayed on the digitized
sky survey image of Cam~B. Consistent with the optical
image the HI isophotes suggest that galaxy is seen at a fairly high 
inclination. Figure~\ref{fig:hisb} shows the deprojected radial
surface density profile of the HI obtained from fitting elliptical
annuli to the HI moment~0 image. A gaussian fit is shown superimposed;
as can be seen the surface density $\Sigma_{HI}(r)$ is well represented 
by a Gaussian:

\begin{equation}
\Sigma_{HI}(r)=\Sigma_0\times e^{-r^2/2r^2_0}
\label{eqn:hisb}
\end{equation}

with $r_0=40.7^{''}\pm1.6^{''}$ (corresponding to a linear scale of 0.43~kpc)
and $\Sigma_0=5.9\pm0.2 M_\odot~pc^{-2}$. 

	The inclination of the HI disk was determined from ellipse fitting 
to the HI isophotes and was found to be $65\pm 5$ degrees (where we have 
assumed an intrinsic thickness ratio $q_0 = 0.25$, see 
Sec.~\ref{sec:massmodels}). The position angle obtained from the ellipse 
fitting was found to be $215\pm5$ degrees. The values of inclination and 
position angle found from the HI distribution are in good agreement
with the values obtained from the optical image (section~\ref{ssec:opt_res}).

\subsection{ HI Kinematics}
\label{ssec:kinematics}

\begin{figure}[h!]
\begin{center}
\epsfig{file=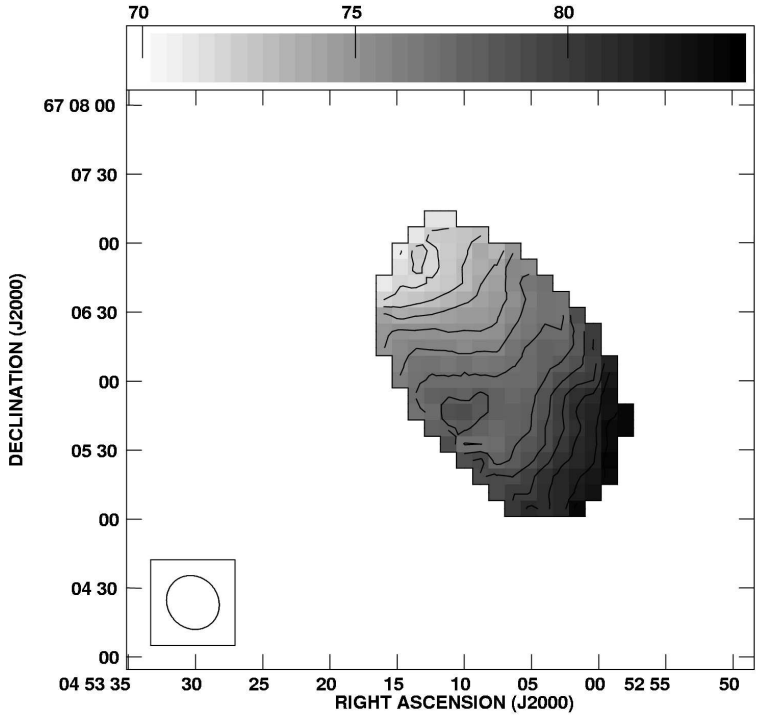,width=4.0in}
\end{center}
\caption{ The HI velocity field of Cam~B at $24^{''}\times 22^{''}$ arcsec
	  resolution. The contours are in steps of 1~km~sec$^{-1}$ and
          range from 70.0~km sec$^{-1}$ (the extreme North West contour)
	  to 84.0~km~sec$^{-1}$ (the extreme South East contour).
         }
\label{fig:mom1}
\end{figure}

	The velocity field  derived from the moment analysis of 
$24^{''}\times22^{''}$ resolution data is shown in Figure~\ref{fig:mom1}. The
velocity field is regular and the isovelocity contours are approximately
parallel, this is the signature of rigid body rotation. The kinematic major 
axis of the galaxy has position angle $\approx{215}^o$, i.e. is aligned with 
the major axis of both the HI distribution (see section~\ref{ssec:hi_flux})
and the optical image (see section~\ref{ssec:opt_res}). Cam~B is the faintest
galaxy known to show such regular kinematics. It is worth noting that
the high velocity resolution ($\sim 1.6$~km~sec$^{1}$) of our observations was
critical in determining the velocity field. Synthesis observations often
use much poorer resolution (eg. \citeasnoun{lo93} used a velocity resolution
of $\sim 6.2$~km~sec$^{-1}$), at these resolutions it would be difficult 
to discern systematic patterns, if any, in the velocity fields of faint
dwarfs.

\section{HI Rotation Curve}
\label{sec:rotcur}

          The rotation curve was derived from the HI velocity field using
the usual tilted ring model (Warner et al. 1973). Three different velocity
fields were used, viz. those derived from the $40^{''}\times38^{''}$,
$24^{''}\times22^{''}$ and the $16^{''}\times14^{''}$ resolution data cubes.
An attempt to derive the rotation curve  leaving the center and inclination
as a free parameters in the fit did not lead to reliable results, consistent
with the solid body like rotation field. The centre of the galaxy  was hence
fixed to be the center of symmetry of the HI moment~0 map and V$_{sys}$ was
fixed at the value of 77.0 km sec$^{-1}$  which was obtained from the single
dish profile.  Again guided by the optical and HI morphology the
inclination of all annuli was fixed at a value of ${65}^o$ and the position
angle at $215^{o}$. The rotation 
curve was then computed using the GIPSY task ROTCUR by breaking up the 
galaxy into annuli (each of width half that of the synthesized beam) and 
fitting to the velocity field in each  annulus keeping all parameters 
except the circular velocity V$_c$ fixed.

\begin{figure}[h!]
\begin{center}
\rotatebox{-90}{\epsfig{file=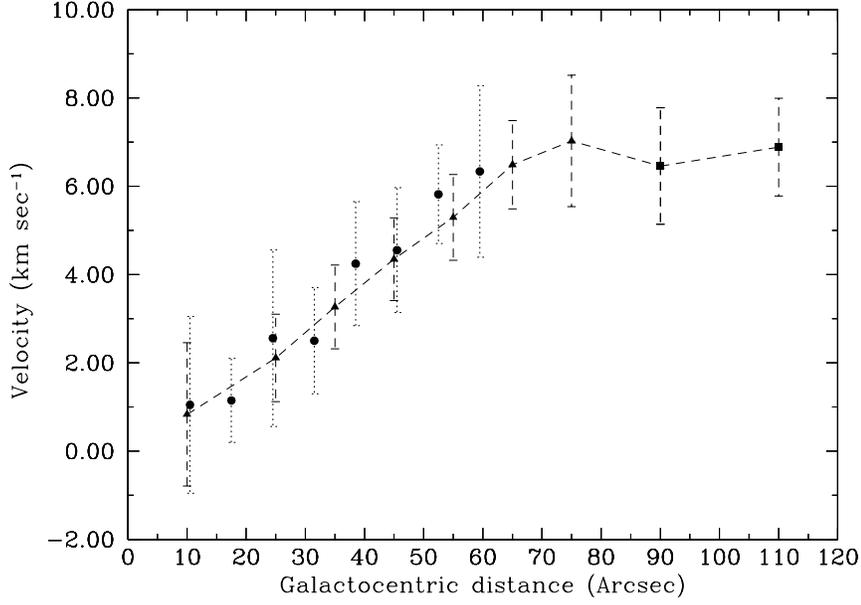,width=3.5in}}
\end{center}
\caption{ Rotation curve derived from the intensity 
	weighted velocity field. Filled circles, triangles and  squares 
        show the rotation curve derived from the $16^{''}\times14^{''}$,
	$24^{''}\times22^{''}$ and $40^{''}\times38^{''}$ resolution
	velocity fields respectively. The dotted line shows the adopted
	rotation curve.
        }
\label{fig:rotcur}
\end{figure}

         The derived rotation curve is shown in Fig.~\ref{fig:rotcur}. The 
rotation curves obtained from $24^{''}\times22^{''}$ and the 
$16^{''}\times14^{''}$ resolution velocity fields agree to within the 
error-bars. For the rest of the analysis, the rotation curve obtained 
from the $24^{''}\times22^{''}$ resolution cube is used  in the inner 
region (up to $80^{''}$) of the galaxy while the outermost points 
(from $80^{''}-120^{''}$) are taken from the lower resolution 
$40^{''}\times38^{''}$ resolution velocity field. From Fig.~\ref{fig:rotcur}
one can see that the maximum  velocity reached by the rotation curve 
($\approx$ 7.0 km sec$^{-1}$) is comparable to the observed velocity 
dispersion i.e $V_{max}/\sigma \approx$1.0.

\begin{figure}[h!]
\begin{center}
\rotatebox{-90}{\epsfig{file=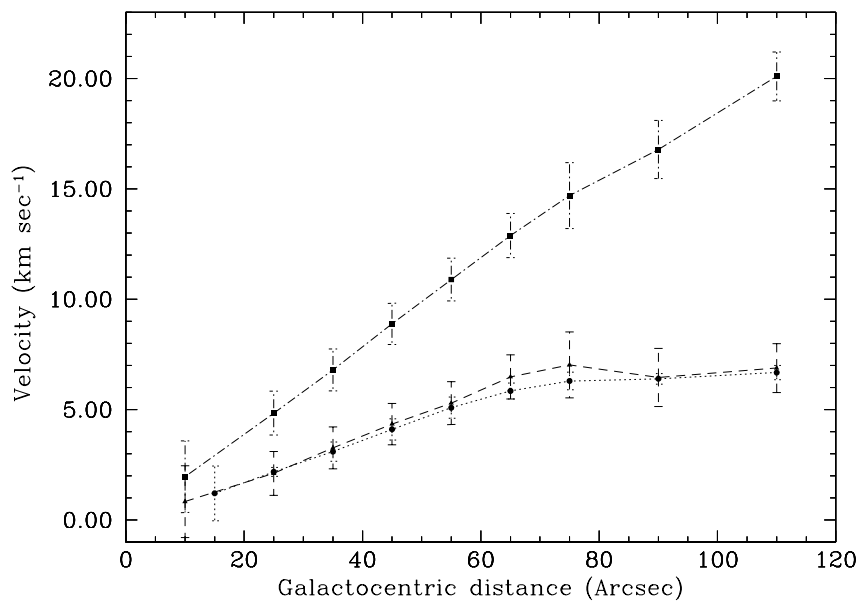,width=3.5in}}
\end{center}
\caption{ The final adopted rotation curve (dashes), the rotation curve
          after beam smearing (dots) and the rotation curve after applying
	  the asymmetric drift correction (dash dots).             
        }
\label{fig:corr_curve}
\end{figure}

	The effect of beam smearing in the derivation of HI rotation curves
has been the subject of recent debate. To estimate the effect of beam
smearing on our data we created a model data cube using the observed 
rotation curve and column density profile and smoothed it to a beam 
size of $24^{''}\times22^{''}$ using the task GALMOD in GIPSY. Moment 
maps were then made of this data cube and a rotation curve was derived 
in exactly the same manner as for the real data. This model rotation 
curve is shown in Fig.~\ref{fig:corr_curve} as a dotted line. As can 
be seen, this agrees very well with the observed rotation curve. This 
suggests that beam smearing has only a small effect for this data set.

           Since for Cam~B the maximum rotation velocity obtained is 
comparable to the velocity dispersion, random motions provide significant 
dynamical support to the disk. Equivalently, the observed circular velocity
significantly underestimates the centripetal force and hence the implied
dynamical mass. The observed rotation velocities hence need to be corrected
before trying to estimate the dynamical mass. This correction, generally 
termed an ``asymmetric drift'' correction is given by 
(eg. Muere et al. 1996; note that the formula below is slightly different
from the one in Muere et al. 1996 since we have used a sech$^2$ profile
in the vertical direction instead of an exponential one)

$v^2_c=v^2_o - r\times{\sigma}^2\bigl[\frac{d}{dr}(\ln{\Sigma_{HI}})+\frac{d}{dr}(\ln{\sigma}^2)-\frac{d}{dr}(\ln{2h_z})\bigr]$,

	where $v_c$ is the true circular velocity, $v_o$ is the observed
rotation velocity, $\sigma$ is the velocity dispersion, and $h_z$ is the
scale height of the disk. Strictly speaking, asymmetric drift corrections
are applicable to collisionless stellar systems for which the magnitude 
of the random motions is much smaller than that of the rotation velocity. 
However, it is often used even for gaseous disks, where the assumption 
being made is that the pressure support can be approximated as the gas
density times the square of the random velocity.  For the case of Cam~B, in 
the absence of any direct measurement of $h_z$ we  assume $d(\ln(h_z))/dr=0$,
(i.e. that the scale height does not change with radius) and also use the 
fact that $\sigma^2$ is constant across the galaxy, to get:
\begin{equation}
v^2_c=v^2_o - r\times{\sigma}^2\bigl[\frac{d}{dr}(\ln{\Sigma_{HI}})\bigr].
\label{eqn:adrift}
\end{equation}
Using the fitted Gaussian profile to the radial surface density distribution,
(see eqn~\ref{eqn:hisb}) we obtain
\begin{equation}
v^2_c=v^2_o + r^2\sigma^2/r^2_0.
\label{eqn:corr_curve}
\end{equation}
   Finally, the observed $\sigma_{obs}$ of 7.3~km$^2$ sec$^{-2}$ (see 
section~\ref{ssec:hi_flux}) needs to be corrected for the finite velocity 
resolution of the observations as well as the contribution of the rotation
velocity gradient over the finite size of the beam. The correction is:

$\sigma_{true}^{2}=\sigma_{obs}^2-\Delta v^2-
	\frac{1}{2}{b}^2{({\nabla}v_o)}^2$,

    where  $\sigma_{true}$ is the true velocity dispersion, $\Delta v$
is the channel width, $b$ characterizes the beam width (i.e. the beam is
assumed to be of form $e^{-x^2/b^2}$) and $v_o$ is the observed 
rotation velocity. This correction turns out to be small; after putting
in the appropriate values in the above equation we get $\sigma^2_{true} 
\approx 49.0$~km$^2$~sec$^{-2}$.  Substituting this value back into 
Eqn.~(\ref{eqn:corr_curve}) to get the final ``asymmetric drift'' 
corrected rotation curve, which is shown as the dash dot line in 
Fig.~\ref{fig:corr_curve}.

\section{ Mass Models}
\label{sec:massmodels}

      In this section we attempt to decompose the mass distribution of
Cam~B into contributions from the stellar disk, the gaseous disk and 
the dark matter (DM) halo. 

	  For the stellar disk we assume a constant (M/$L_V$) ratio
$\Upsilon_V$ with the radius. Recall from section~\ref{ssec:opt_res} 
that the galaxy color was approximately constant in the inner regions 
(within $\sim 35^{''}$), but there is a gradient in the color in the 
outer parts. This suggests that there could be a gradient in mass to 
light ratio in the outer parts of the galaxy. However, since it turns
out that the stellar disk makes a negligible contribution to the total
mass in the outer parts of the galaxy, the assumption of a constant 
$\Upsilon_V$ throughout the galaxy will not lead to substantial 
uncertainty in the mass of the dark halo. Similarly, the deviations 
from the exponential fit in the outer parts of the galaxy (which are 
related to the very faint extensions to the north in Fig.~\ref{fig:overlay})
has only a small effect on the fitted mass model. We hence take the 
stellar disk to be an  exponential  disk (see section~\ref{ssec:opt_res}) 
with an intrinsic thickness ratio ($q_o$) of 0.25 (see below). We further
assume that the density distribution in the vertical($z$), direction 
falls off like sech$^2(z/z_0)$ with $z_0$ independent of galacto-centric
radius. This is a reasonable approximation for disk galaxies 
\citeaffixed{kruit81,grijs97}{see eg.}. In the absence of any  prior 
knowledge of $\Upsilon_V$, it is taken to be a free parameter during 
the modeling.

           The contribution of the gaseous disk to the observed rotation 
curve is calculated using HI surface mass density  profile given in 
Fig.~\ref{fig:hisb}.  The contribution of primordial Helium is taken 
into account by multiplying HI densities by a factor of 1.4. There is
little evidence that dwarf galaxies contain substantial amounts
of molecular gas \citeaffixed{israel95,taylor98}{see eg.},
so no correction has been made for molecular gas. We also neglect the
contribution of ionized gas, if any. Not much is  known about the vertical 
distribution of gas in disk galaxies, however there is some evidence
for similar  vertical distributions of the HI and stellar disk 
\citeaffixed{bottema86}{eg.}. For the intrinsic thickness ratio of the 
gaseous disk we hence assume  $q_0=0.25$ and also assume that the shape of the profile 
in the vertical direction is given by sech$^2(z/z_0)$. To check the validity of
these assumptions we tried modeling the HI distribution and velocity 
field with various vertical profiles and intrinsic thickness ratios using 
the GIPSY task GALMOD. It was found that the best (visual) match between
the model  and the observed maps were obtained with a 
sech$^2(z/z_0)$ profile and  an intrinsic thickness ratio of 0.25. The
geometries of all the disk components being thus fixed, the circular 
velocities of the disk components were computed using the formulae given by 
\citeasnoun{casertano83}.

                For the dark matter we considered two types of density 
profiles, a modified isothermal halo and an NFW halo. The modified isothermal
halo has a density profile given by:

$\rho_{iso}(r)=\rho_0[1+{(r/r_c)}^2]^{-1}$,

where, $\rho_0$ is the central density of the halo and 	$r_c$ is the core radius.
The corresponding circular velocity is given by

$v(R)=\sqrt{4\pi G \rho_0 r^2_C\bigl[1-\frac{r}{r_c}\tan^{-1}(\frac{r}{r_c}\bigr)]}$.

  The NFW halo density is given by

$\rho_{NFW}(r)=\rho_i/[(r/r_s)(1+r/r_s)^2]$,

  where, $r_s$ is the characteristic radius of the halo and $\rho_i$ is 
the characteristic density. The circular velocity can be written as:

$v(R)=v_{200} \sqrt{\frac{\ln(1+cx)-cx/(1+cx)}{x[\ln(1+c)-c/(1+c)]}}$,

where,  $c = r_{200}/r_s$, $x = r/r_{200}$; $r_{200}$ is the distance at which the mean
density of the halo is equal to 200 times the critical density and 
$v_{200}$ is the circular velocity at this  radius.

	We consider first fits using a modified isothermal halo.
Because the rotation curve (Fig.~\ref{fig:corr_curve}) is rising till the
last measured point,  the parameter $r_c$ cannot be constrained. (To
constrain $r_c$ one needs the rotation curve to transition from rising
linearly to being more or less flat, as is typically seen for spiral
galaxies). We are therefore left with two free parameters, viz., the
mass to light ratio of the stellar disk, $\Upsilon_V$, and the core
density of the halo $\rho_0$. The modeling procedure consists of a $\chi^2$
minimization of 

\begin{figure}[h!]
\begin{center}
\rotatebox{-90}{\epsfig{file=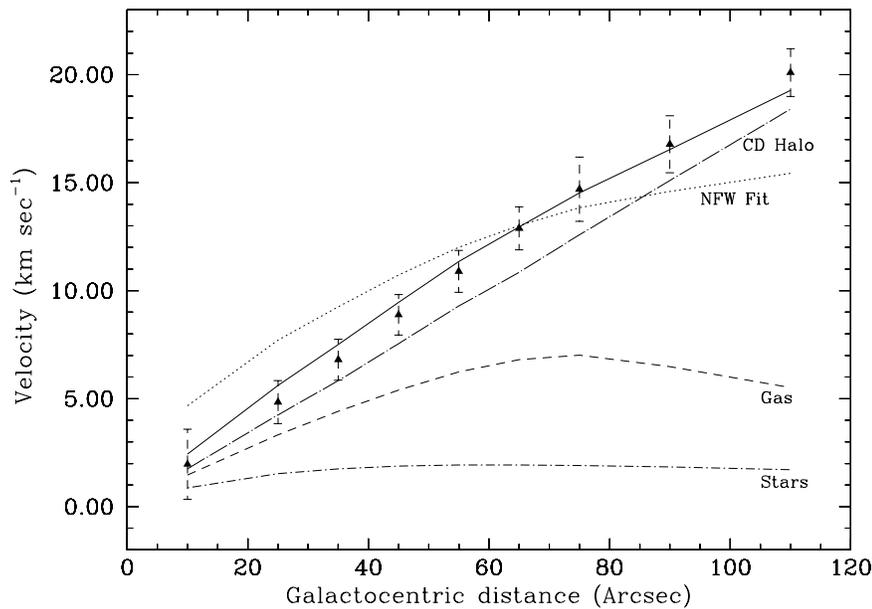,width=3.5in}}
\end{center}
\caption{Mass models for \astrobj{Cam B} using the corrected rotation curve.
The points are the observed data. The total mass of gaseous disk (dashed line)
is $6.6\times10^6 M_\odot$.The stellar disk (short dash dot line) has
$\Upsilon_V=0.2$, giving a stellar mass of $0.7 \times10^6 M_\odot$. The
best fit total rotation curve for the constant density halo model is shown as
a solid line, while the contribution of the halo itself is shown as a
long dash dot line (the halo density is density $\rho_0=13.7\times10^{-3}
M_\odot$ pc$^{-3}$). The best fit total rotation curve for an NFW type halo
(for $c=1.0$ and $\Upsilon_V=0.0$) is shown as a dotted line. See the
text for more details.}
\label{fig:mass}
\end{figure}

 $v^2_c-\Upsilon_V v^2_{*} - v^2_{g} - v^2_{h}(\rho_0)$,

  where,  $v_c$ is  the corrected rotation velocity (see 
eqn.~\ref{eqn:corr_curve}),  $v_{*}$, $v_g$, and $v_h$ are the circular
speeds of the stellar disk,  gaseous disk  and  the dark halo respectively.
This minimization was done using the GIPSY task ROTMAS.
Since there are heuristics involved in computing the error bars
on $v_c$, it is not possible to rigorously translate the minimum 
$\chi^2$ value into a confidence interval for the parameters of the
fit. However, a lower $\chi^2$ value does imply a better fit 
\citeaffixed{bosch01}{see also the discussion in}. We found that 
the $\chi^2$ continuously decreases as $\Upsilon_V$ decreases.  
The observed  (B-V) for Cam~B is 0.8, which corresponds to 
$\Upsilon_V$ is $\sim 1.8$ \citeaffixed{bell01}{from the low metalicity 
Bruzual \& Charlot SPS model using a modified Salpeter IMF,}. Using this
value of $\Upsilon_V$  gives $\rho_0=12.0\times 10^{-3}~M_\odot 
~pc^{-3}$ and a reduced $\chi^2$ of 1.9. For comparison, fixing
$\Upsilon_V$ to be 1.0, gives a reduced $\chi^2$=0.96 and a
best fitting $\rho_0=12.4\times 10^{-3}M_\odot$pc$^{-3}$. 
At the two extremes, if we assume $\Upsilon_V = 0$ we get 
$\rho_0=14.0\times 10^{-3} M_\odot$pc$^{-3}$ (and a reduced 
$\chi^2$=0.2) and  $\Upsilon_V = 2$ (which substantially overpredicts 
the observed rotation curve at small radii, and is hence an upper 
limit to $\Upsilon_V$) we get $\rho_0=11.0\times 10^{-3} 
M_\odot$pc$^{-3}$  (and a reduced $\chi^2$=2.2). As can be seen 
the central halo density $\rho_0$ is relatively insensitive to 
the assumed $\Upsilon_V$, and is well determined.  To illustrate 
the general quality of fit we show in Fig.~\ref{fig:mass} the 
mass decomposition for a modified isothermal halo using 
$\Upsilon_V =0.2$.

      A similar procedure was tried using a dark matter halo of the
NFW type, but no reasonable fit could be obtained for any value of
$\Upsilon_V$. Essentially \astrobj{Cam B} is dark matter dominated and has a
linear rotation curve while the NFW halo provides a poor fit for
linear rotation curves. As an illustration we show in Fig.~\ref{fig:mass}
the best fit rotation curve for an NFW halo with $c=1.0$, $\Upsilon_V=0.0$,
and $v_{200}$ chosen to minimize the $\chi^2$. These are already unphysical
values for these parameters, and increasing either $c$ or $\Upsilon_V$ only
worsens the quality of the fit.

\section{Discussion}
\label{sec:dis}

   The mass of the gas disk in Cam~B is $M_{gas}$=$6.6\times10^6 M_\odot$
and the adopted ($\Upsilon_V=1$) model gives  the mass of stellar disk to 
be $M_*$=$3.5 \times 10^6 M_\odot$. From the last measured 
point of the  observed rotation curve, we get a total dynamical mass of 
$M_T$=1.1$\times 10^8 M_\odot$, i.e. at the last measured point 
more than 90\% of the mass  of Cam~B is dark. Like other faint dwarf 
galaxies, Cam~B is also dark matter dominated in the inner regions, by 
the time one reaches a galacto-centric radius of two (optical) disk 
scale lengths, the mass of the stellar and gas disks is small compared 
to the mass in the dark matter.
    
\begin{figure}[t!]
\begin{center}
\epsfig{file=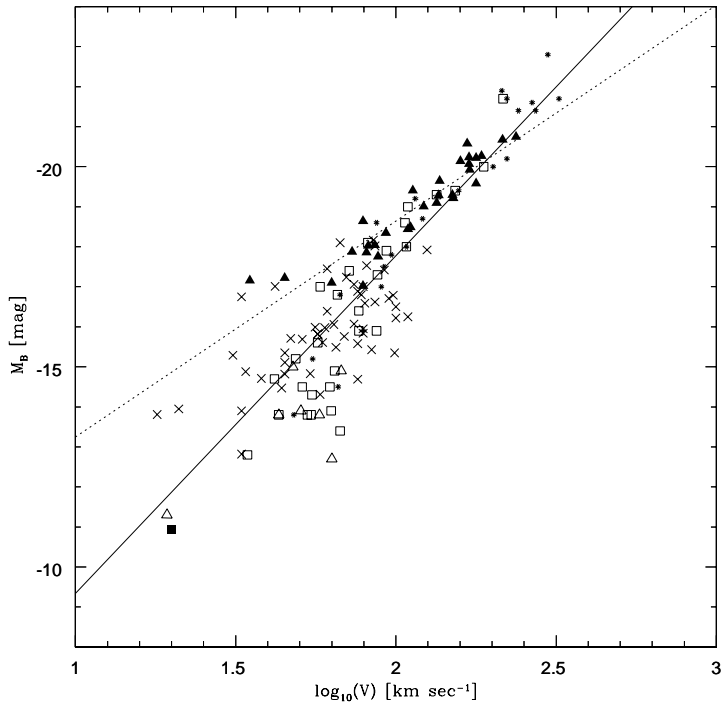,width=4.0in}
\end{center}
\caption{ Absolute B~magnitude vs the log of the maximum rotation velocity.
  	  Stars are galaxies from Broeils~(1992), empty triangles
	  galaxies from C\^{o}t\'{e} et al.~(2000),  filled triangles 
	  from Verheijen~(1997), empty squares from Stil~(1999) and 
	  crosses from Swaters~(1999). Cam~B is shown by a filled hexagon
	  (where the ``assymetric drift'' corrected velocity is used) and 
	  an empty hexagon (where the observed peak velocity is used). The 
	  dotted line is the best fit to the galaxies from the 
	  Verheijen~(1997) sample alone, while the solid line is the
	  best fit to the entire sample.
	} 
\label{fig:TF}
\end{figure}

       In Fig.~\ref{fig:TF}  the  maximum rotation velocity for Cam~B (after
correction for ``asymmetric drift'') is plotted against the absolute blue
magnitude. The same quantities for several other spiral and dwarf galaxies
with measured HI rotation curves are also shown. The samples from which
these galaxies have been drawn are listed in the figure caption. A clear 
trend is seen between the maximum velocity and absolute magnitude (which is 
of course the basis of Tully-Fisher relationship). We are forced to use 
the blue magnitude (even though it is in general a poor quantity for this
purpose, see below) because it is the only band  for which measurements 
exist for both Cam~B as well as all the other samples that we have used. 
Primarily because of uncertain and absolute magnitude dependent internal
extinction corrections, the B~band 
Tully-Fisher relationship has been the subject of recent debate. 
\citeasnoun{stil99} noted that dwarf galaxies lie systematically below
the Tully-Fisher relationship defined by bright galaxies (the dotted line
in Fig.~\ref{fig:TF}) i.e. dwarfs are comparatively underluminous for
their given circular velocity. \citeasnoun{swaters99} noted the same
effect for the R~band Tully-Fisher relationship and suggested that
the problem could be ameliorated by applying a ``baryonic correction''
i.e. a correction for the fact that dwarf galaxies have still to convert
much of their gas into stars. \citeasnoun{pierini99} found that the 
adopted correction for internal extinction makes a substantial 
difference to the slope of the B~band Tully-Fisher relationship,
and that depending on the internal extinction correction applied one 
could reduce (but not completely eliminate) the need for a baryonic
correction.  Finally, \citeasnoun{mcgaugh00} state that due to the 
problem of internal extinction the blue band Tully-Fisher relation 
actually gives no clear evidence for the need for  a baryonic correction, 
but that the need for such a correction is clear when one uses near
infrared absolute magnitudes (i.e. bands which are little affected
by internal extinction). 

	The galaxy magnitudes in Fig.~\ref{fig:TF} range from $M_B \sim 
 -23$ to $M_B\sim -11$, and we also show in the figure the best fit 
straight line (the solid line) between M$_B$ and $V_{max}$ over this
entire magnitude range. Cam~B lies close to this line if one uses the 
``asymmetric drift'' corrected rotation velocity.  The scatter about 
the best fit line is $\sim 1$~magnitude, a large part of this is probably
due to (i)~uncertainties in the distances to the galaxies, and 
(ii)~the  different prescriptions for correcting for
internal extinction that have been applied to the different samples.
As expected from the discussion above, one can see that galaxies with 
large velocity widths lie systematically above this best fit slope, 
while galaxies with low velocity widths lie systematically below this
slope (with the break occurring at $\sim 100$ km~sec$^{-1}$ -- this
corresponds well to the break velocity seen in the near infrared 
Tully-Fisher diagram, \cite{mcgaugh00}). 

	Finally we note that Fig.~\ref{fig:TF} does not show any 
particular flattening of the slope around M$_B$=$-$14.0 (as claimed  
by \citeasnoun{cote00}). Part of the reason for this difference is 
probably that \citeasnoun{cote00} used rotation velocities from 
the \citeasnoun{lo93} sample, while we have not. The observations
of \cite{lo93} used a velocity resolution of $\sim 6.5$~km~sec$^{-1}$,
and also often lacked sensitivity to the low extended HI distribution,
for both of these reasons the V$_{max}$ of the galaxies in this sample
could be underestimated \citeaffixed{skillman96,hoffman96}{see eg.}. 

\begin{figure}[t!]
\begin{center}
\epsfig{file=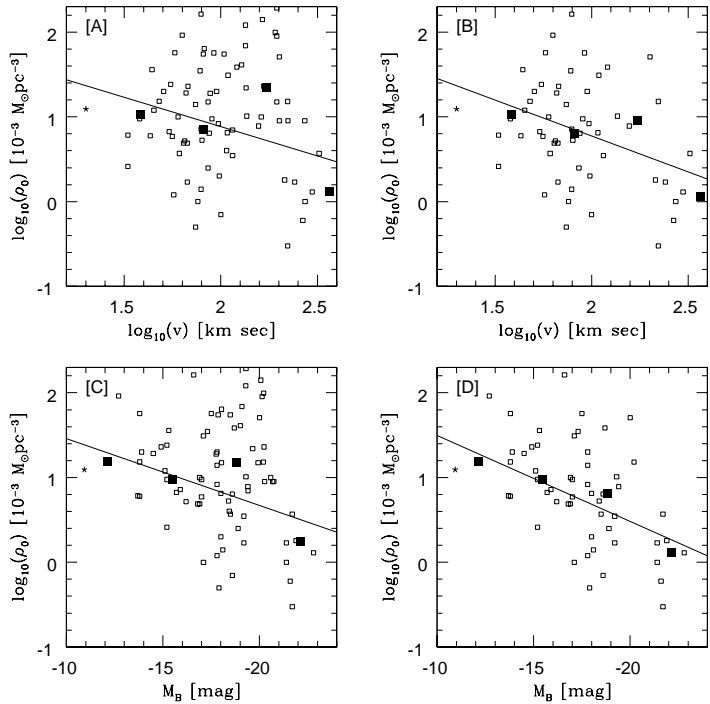,width=4.0in}
\end{center}
\caption{ Scatter plots of the central halo density against the
	  circular velocity (panels [A] and [B]) and the absolute
	  blue magnitude (panels [C] and [D]). The data (empty squares)
	  are from Verheijen~(1997), Broeils~(1992), 
	  C\^{o}t\'{e} et al.~(2000) and Swaters~(1999). The filled 
	  squares are the medians of the binned data, and the straight 
	  lines are the best fits to the binned data. Cam~B is shown 
	  as a star. }
\label{fig:dens}
\end{figure}

	In Sec.~\ref{sec:intro} we had discussed that based on galaxy
formation simulations, several authors have suggested that smaller
galaxies should in general have a higher central halo density than
bigger galaxies. We plot in Fig.~\ref{fig:dens} the halo density
of galaxies (drawn from the same samples as for Fig.~\ref{fig:TF},
except that the sample from \cite{stil99} had to be dropped since
mass decompositions were not available for this sample)
as a function of their circular velocity (panels [A] \& [B])
and as a function of the absolute blue magnitude (panels [C] \& [D]). 
From the raw data itself one finds in general no clear correlation
between halo density and blue magnitude or circular velocity. To 
see if this the effect of a large scatter, we also bin 
the data in 4 bins and show the median halo density in each bin 
(the soild points). As can be seen the binned data do show a trend
of increasing halo density with decreasing blue luminosity or 
circular velocity. Interestingly, the correlation is sharper when
the Ursa Major sample of \cite{verheijen97} is excluded (panels [B] \& [D]). 
If this is a real effect, then it is perhaps related to the higher mean 
galaxy density of the Ursa Major sample. The solid lines shown in
panels~[B] \& [D] are $\rho_0 \propto V_{max}^{-0.85}$ and 
$\rho_0 \propto M_{B}^{0.25}$ respectively. We also note that the 
halo densities being plotted have been computed for ``maximum disk'' 
modified isothermal halo fits, and that the correlation we find is 
related to (but not identical to) the statement often made in the
literature that the in centers of large galaxies the baryonic mass
is more dominant than in the centers of dwarf galaxies.
    
	To conclude, we have presented optical images and HI 21cm
synthesis data for the faint dwarf galaxy Cam~B. We find that 
Cam~B has a very regular velocity field that can be well fit with
the standard tilted ring model, despite the fact that its observed
peak rotation velocity is of the same order of magnitude as the
random motions of the gas. The very high velocity resolution of 
our observations ($\sim 1.6$ km sec$^{-1}$) were crucial in order
to determine the systematic patterns in the velocity field of the
HI disk. Mass modeling of the galaxy shows that its rotation curve
can be well fit by a constant density halo, with density 
$\sim 12 \times 10^{-3}M_\odot$pc$^{-3}$. This measurement of the
halo central density is not very sensitive to the assumed mass to
light ratio ($\Upsilon_V$) of the stellar disk. Like other faint
dwarf galaxies the mass of Cam~B is dominated by the dark matter
halo; at the last measured point $>90\%$ of the mass is in the dark
halo. The dominance of the dark matter halo together with the 
linear shape of the rotation curve (after correction for 
``asymmetric drift'') mean that one cannot obtain a good fit 
to the rotation curve using a NFW halo regardless of the 
assumed $\Upsilon_V$. Finally, combining the data for Cam~B with
the data available in the published literature for other dwarf
and normal disk galaxies, we find a weak trend for the
central halo density to decrease with increasing galaxy size.

{\bf Acknowledgements:}
	These observations would not have been possible without the
many years of dedicated effort put in by the GMRT staff in order to build 
the telescope.  The GMRT is operated by the National Centre for Radio 
Astrophysics of the Tata Institute of Fundamental Research.
We are grateful to D. J. Bomans for generously sharing his data 
and results prior to publication.

\end{document}